# Hidden momentum in a moving capacitor

Giovanni Asti

*Università di Parma, Dipartimento di Fisica e Scienze della Terra*

*43100 Parma, Italy*

email: *giovanni.asti@unipr.it*

A very simple system like a parallel-plate capacitor reveals striking features when we examine the peculiar phenomena appearing when it is moving at low speed in different directions. Both hidden momentum and hidden energy appear and their addition, with their sign, to the corresponding electromagnetic component results in the expected ordinary kinetic momentum or energy of the electrostatic mass equivalent. What's happening is that passing from one inertial reference frame to another, part of the energy or momentum is transferred from the electromagnetic component to the material part of the system or the other way around. A paradoxical self-accelerating behavior is evidenced if one admits that the capacitor is discharging through an electrical resistance during its motion. It is shown that one must take into account the mass of the produced heat.

## I. INTRODUCTION

It is instructive for the student of electromagnetism to deeply examine the complex and sometimes surprising interplay of fields, stresses, momentum and energy when dealing with moving bodies. This happens even for very simple systems such as a capacitor, or a solenoid, because electromagnetism is intrinsically bonded with special relativity and unique in revealing the typical clamorous relativistic effects even at extremely low speeds. The student will appreciate the fact that perfect consistency between the view from different inertial systems is only achieved when length contraction and non simultaneity effects are admitted. On the other hand it is known that this fact was at the basis of the well known paradoxes in electromagnetism at the beginning of the past century. Consider the title itself of the fundamental article where Albert Einstein presents his theory of relativity ("Zur Electrodynamik bewegter Korper", 195)[1]. Of particular interest is the necessity of a hidden momentum and energy in the moving body if fundamental conservation laws should not be broken. The hidden momentum is the consequence of the fact that an energy current S should imply a corresponding momentum density $g=S/c^2$. That this relation should have universal validity was already supposed by Planck [2]. In fact it is at the basis of the explanation of the negative result in the historical Trouton-Noble experiment in 1901 [3]. The presence of hidden momentum has been discussed in the literature particularly in the case of stationary systems in the presence of current carrying bodies and /or magnetic fields and charges. The presence of hidden momentum is necessary in order to guarantee that the stationary system have zero total momentum. [4, 7]. Hidden momentum has been analyzed in general in moving system by Comay [8]. Other authors have treated the case of a moving capacitor [9-11] from different points of view.

In the present work a very simple system is chosen, namely a plane electric capacitor: it is shown that a variety of peculiar effects are obtained simply because it is put in motion at uniform velocity *v*. In particular comparison is made of electromagnetic momentum and energy with hidden momentum and energy. Both cases of *v* perpendicular and parallel to the field will be discussed showing that different fractions of mass-energy conversion are implied in the different configurations.

## II. CAPACITOR MOVING IN A DIRECTION PARALLEL TO THE PLATES
### A. General considerations

Let us consider a charged plane capacitor at rest in an orthogonal reference frame $S$ (laboratory frame). It has rectangular shape with side $a$ parallel to $x$-axis and the other side parallel to $y$-axis with length $l$. The two metal plates have negligible thickness and are separated by the displacement vector $\mathbf{s}=\mathbf{p}/q$, the ratio of the dipole vector $\mathbf{p}$ to the modulus of the electric charge $q$. The reference frame $S'$ is translating at a velocity $v$ in the direction of the x-axis. The spatial axes of system $S'$ preserve the orientations of those of system $S$. Hence the capacitor is moving at a velocity $v'=-v$ with respect to frame $S'$. We want to determine the linear hidden momentum and energy from the point of view of observer $O'$[Note 1]*. In particular the hidden momentum of the two plates can be determined once we have calculated the mechanical tension caused by the mutual repulsion among the electric charges lying on the plates. Hence the edges of the plates are subjected to a surface tension $\tau$ which can be obtained supposing to extend by an amount $\delta x$ the side $a$: the charge is invariant, the thickness remains the same while the surface of the plate, $A=al$, increases by an amount $\delta A = l\delta x$ [note 2]**. The change of the electrostatic energy is

$$\delta E = \frac{q^2 s}{2\varepsilon_0}\left(\frac{1}{A+l\delta x} - \frac{1}{A}\right) = \frac{q^2 s}{2\varepsilon_0}\delta A \frac{\partial}{\partial A}\left(\frac{1}{A}\right) = -\frac{q^2 s}{2\varepsilon_0}\frac{\delta A}{A^2} = -\frac{\sigma^2 s}{2\varepsilon_0}l\delta x \qquad (1)$$

where $\sigma$ is the surface charge density of the positive plate. Energy conservation requires that this energy change should be opposite to the mechanical work made by the above mentioned surface tension

$$\delta E = -\frac{\sigma^2 s}{2\varepsilon_0}l\delta x = -2l\tau\delta x \qquad (2)$$

from which we obtain

$$\tau = +\frac{\sigma^2 s}{4\varepsilon_0} \qquad (3)$$

The plus sign indicates that it is not obviously a surface tension similar to that of a soap bubble but on the contrary an outward directed tension tending to expand the surface. Note that the obtained average lateral pressure is $2\tau/s = w$, the electrostatic energy density, as expected.

Thus we can calculate the hidden momentum inside each plate of the capacitor. The energy current that $O'$ evaluates is

$$G = -2lv'\tau = 2lv\tau = 2\frac{sw}{2}lv = wslv = \frac{\sigma^2 slv}{2\varepsilon_0} \qquad (4)$$

directed in the opposite direction with respect to the motion of the capacitor because the tension on the forward side is making a positive work, thus transferring energy to the back side. The hidden momentum $p_h$ inside the capacitor is the energy current divided by $c^2$ and multiplied by $a$, the longitudinal dimension:

$$p_h = wslva/c^2 = wsAv/c^2 = \frac{\mu_0}{2}\sigma^2 sAv. \qquad (5)$$

$p_h$ refers to the whole capacitor, so that it includes the contribution of both plates [note 3]***. As is evident from (5), $p_h$ turns out to be counterbalanced by what we could call the "kinetic" linear momentum, i.e. the one associated with the translation of the electric field confined inside the capacitor. It is obtained by determining the mass of the electrostatic energy and multiplying by $v$ [Note 1]*:

$$p_e = m_e v' \quad \text{with} \quad m_e = \gamma w s A / c^2 \cong w s A / c^2 \tag{6}$$

So we have

$$p_e = -p_h \tag{7}$$

But obviously we cannot ignore the role of the electromagnetic field itself. In fact during their motion the two plates are swept by opposite surface currents having density

$$\vec{k}_+ = \sigma \vec{v}' = -\sigma \vec{v} \quad \text{and} \quad \vec{k}_- = \sigma \vec{v} \tag{8}$$

generating a magnetic induction of intensity

$$\vec{B} = \mu_0 \vec{k}_+ \times \left(-\frac{\vec{s}}{s}\right) \tag{9}$$

inside the capacitor. Hence the linear momentum turns out to be

$$\vec{q}_{em} = sA |\vec{D} \times \vec{B}| = \left| -\mu_0 s A \sigma \frac{\vec{s}}{s} \times \left[ \vec{k}_+ \times \left(-\frac{\vec{s}}{s}\right) \right] \right| = -\mu_0 A \sigma \frac{\vec{s}}{s} \times (\sigma \vec{v} \times \vec{s}) = -sA \mu_0 \sigma^2 \vec{v} \tag{10}$$

Note that $q_{em}$ has the same sign of the velocity of the capacitor, $v' = -v$, and

$$q_{em} = 2 p_e = -2 p_h. \tag{11}$$

Finally we recognize that the obtained values of the three kinds of linear momentum are consistently related by the balance equation

$$p_e = q_{em} + p_h \tag{12}$$

**B. Discharging capacitor**

In order to assess the above considerations in a realistic situation let us examine the following example, which could also be presented as an apparent electromagnetic paradox. It is the case of a discharging capacitor in motion. It is immediate to recognize that the the motion couples to electro-dynamical parameters of the system to produce accelerating forces: but the fact is that the capacitor is not certainly accelerating!

So the observer in $S'$ sees the translating capacitor while it is decreasing its charge through the dielectric due to a loss current density

$$\vec{j} = \dot{\sigma} \frac{\vec{s}}{s} \tag{13}$$

This current, together with field $\vec{B}$ given by eq. (9), gives rise to an accelerating force

$$\vec{F}_1 = As\vec{j} \times \vec{B} = A\dot{\sigma}\vec{s} \times \left[\mu_0 \vec{k}_+ \times \left(-\frac{\vec{s}}{s}\right)\right] = -A\mu_0\dot{\sigma}\frac{\vec{s}}{s} \times (\vec{k}_+ \times \vec{s}) =$$
$$-A\mu_0\dot{\sigma}\frac{1}{s}(\vec{k}_+ s^2 - \vec{s}(\vec{k}_+ \cdot \vec{s})) = \mu_0 sA\dot{\sigma}\sigma\vec{v} \qquad (14)$$

which is perfectly balanced by the decrease of the hidden momentum. In fact its time derivative turns out to be

$$\frac{d}{dt} p_h = \frac{1}{2}\mu_0 slav \frac{d}{dt}(\sigma^2) = \mu_0 sA\dot{\sigma}\sigma v \qquad (15)$$

But there is a further force $F_2$ acting on the charges on the plates. It is caused by the induced electric field $E_i$ produced by the changing magnetic flux which crosses the space between them:

$$\phi(B) = sa\vec{j} \cdot \vec{B} = sa\hat{j} \cdot \mu_0 \left[\frac{\vec{s}}{s} \times \vec{k}_+\right] = \mu_0 a\hat{j} \cdot (\vec{s} \times \sigma\vec{v}') = \mu_0 as\sigma v' \qquad (16)$$

Hence, neglecting $s$ with respect to $a$, we have

$$E_i \cong -\frac{d\phi(B)}{dt}/(2a) = -\frac{d(\mu_0 \sigma v' sa)}{dt}/(2a) = \mu_0 \dot{\sigma} vs/2 \qquad (16')$$

and the force acting on the whole charge of the capacitor turns out to be

$$\vec{F}_2 = l \oint \sigma da \vec{E}_i \cong 2A\sigma\mu_0\dot{\sigma}\vec{v}s/2 = \mu_0 sA\sigma\dot{\sigma}\vec{v} = \vec{F}_1 \qquad (17)$$

So $F_2$ is likewise tending to accelerate the capacitor. On the other hand this behavior is also clear from Lenz law. In fact this law requires, with decreasing $\sigma$, and so $k$, an induced electric field $E_i$ tending to maintain $k$. Hence $qE_i$ is always parallel to $v'$ on both plates. This force as a matter of fact does not accelerate the capacitor because it is entirely devoted to provide the necessary impulse to the increasing mass of the heat generated inside the capacitor by the loss current. The heat is indeed produced at the expenses of the electrostatic energy of the capacitor. So the outgoing energy flux caused by the loss current is

$$-\frac{d}{dt}\left(\frac{sA\sigma^2}{2\varepsilon_0}\right) = -sA\sigma\dot{\sigma}/\varepsilon_0 \qquad (18)$$

Hence we have a mass flow entering the material structure of the capacitor, in the form of heat energy, given by

$$\frac{dm_{th}}{dt} = -\frac{sA\sigma\dot{\sigma}}{\varepsilon_0 c^2} = -\mu_0 sA\sigma\dot{\sigma} \qquad (19)$$

As a consequence, in order to keep constant the velocity $v'$, a force is necessary and this force is exactly $F_2$:

$$F = \frac{dp}{dt} = \frac{d(m_{th}v')}{dt} = \mu_0 sA\sigma\dot{\sigma}v = F_2 \qquad (20)$$

Beside this one can easily verify that the total linear impulse is preserved because $q_{em}$ and $p_h$ decrease with time by the same proportion and we see from eq. (12) that $p_e$ will decrease at the same rate, but the fact is that the capacitor will maintain the same velocity v' and the same overall momentum. The reason is that the diminution of the electric mass is exactly compensated by the gain in the material mass of the capacitor in terms of thermal energy.

Note that in all the above considerations we have neglected second order relativistic effects, such as length contraction, because the relations concerning linear momentum we have examined are effects of first order in *v/c*.

**C. Energy relations**

Let us consider now the energy associated with the electromagnetic field as well as the hidden energy of the system. Now we must obviously take into account effects of the higher order in *v/c* and hereafter we will neglect corrections superior to this order. In the *S* reference frame we have only electrostatic energy

$$\mathsf{E}_{em} = \frac{1}{2} sA \frac{D^2}{\varepsilon_0} \qquad (21)$$

It represents a rest mass of electric field given by (compare with eq. 6)

$$m_{0e} = \frac{1}{2} sA \frac{D^2}{\varepsilon_0 c^2} = \frac{1}{2} \mu_0 sAD^2 = \frac{1}{2} \mu_0 sA\sigma^2 \qquad (22)$$

Thus we can think that, from the point of view of the laboratory frame *S*, when the capacitor is in motion, it contributes to the kinetic energy of the system by an amount

$$T_e = \frac{1}{2} m_{0e} v^2 = \frac{1}{4} \mu_0 sA v^2 \sigma^2 \qquad (23)$$

In the reference frame *S'* the electromagnetic energy is given by

$$\mathsf{E}'_{em} = \frac{1}{2} s'A' \frac{D'^2}{\varepsilon_0} + \frac{1}{2} s'A' \frac{B'^2}{\mu_0} \qquad (24)$$

where

$$D' = \gamma D, \quad B' = \mu_0 \gamma \sigma v, \quad A' = A/\gamma, \quad s' = s \qquad (25)$$

with

$$\gamma = \left(1 - \frac{v^2}{c^2}\right)^{-1/2} \qquad (26)$$

We are interested in the variation of the electromagnetic energy due to motion, hence we want to evaluate

$$\Delta \mathsf{E}_{em} = \mathsf{E}'_{em} - \mathsf{E}_{em} = (\gamma - 1)\frac{1}{2}sA\frac{D^2}{\varepsilon_0} + \frac{1}{2}\gamma\mu_0 sA\sigma^2 v^2 = [(\gamma-1)c^2 + \gamma v^2]\frac{1}{2}\mu_0 sA\sigma^2 = \quad (27)$$

$$\left[(1 + \frac{1}{2}\frac{v^2}{c^2} - 1)c^2 + v^2 + ...\right]m_{0e} \cong \frac{3}{2}m_{0e}v^2 = 3T_e = \frac{3}{4}\mu_0 sA\sigma^2 v^2$$

Of this energy variation the magnetic part is

$$\mathsf{E}_m = \frac{1}{2}\gamma\mu_0 sA v^2 \sigma^2 \cong \frac{1}{2}\mu_0 sA v^2 \sigma^2 = m_{0e}v^2 = 2T_e \quad (27')$$

while the remaining part, equal to $T_e$, is the mere variation of the electric part due to contraction of the length l.

Utilizing the principle of relativity of simultaneity used in eq. (5') [note 3] *** we can also calculate the hidden energy. The positive work made by the mechanical reaction forces on the back side is

$$\Delta L_h = 2l\tau \cdot v\Delta t = 2l\frac{\sigma^2 s}{4\varepsilon_0} \cdot \frac{v^2 a}{c^2} = \frac{\mu_0}{2}sA\sigma^2 v^2 = 2T_e \quad (28)$$

So that the hidden energy is $\mathsf{E}_h = -2T_e$. As a result the following energy balance equation occurs between the three terms

$$T_e = \Delta \mathsf{E}_{em} + \mathsf{E}_h \quad (29)$$

which is the corresponding of eq. (12) relative to the linear momentum terms. We see that the kinetic energy of the electric field of the capacitor is divided in a positive part $3T_e$ in the electromagnetic field and a negative part $-2T_e$ inside the matter that constitutes the capacitor. It means that during the acceleration process a conversion of mass to energy occurs. This result is consistent with what in parallel occurs to the linear momentum, but with different numerical factors. The different numerical factors are connected with the fact that the energy-momentum relation is different: It is

$\mathsf{E} = pv/2$ and $\mathsf{E} = pv$ for the kinetic and the hidden term respectively.

Further consideration can be done taking into account the role of the forces acting on the electric charges during the acceleration process of the system. In fact we can calculate the impulse provided by the external accelerating force. It is opposite to that due to the induced electric field $E_i$ caused by the increasing intensity of the magnetic field:

$$\oint E_i dl = -\frac{d}{dt}\varphi(B), \quad (30)$$

from which we have $\quad E_i \cong -\frac{1}{2}\mu_0 \sigma s \dot{v}'.$ $\quad (31)$

Then the impulse is

$$p_q = -\int 2qE_i dt = \int \mu_0 \sigma^2 \dot{v}' sA dt = \mu_0 \sigma^2 v' sA = q_{em}, \quad (32)$$

as expected. In fact the impulse received by the charges is realized in the linear momentum of the electromagnetic field. However one half of this momentum is balanced by the hidden momentum.

What remains is properly the other half, the kinetic momentum (eq.s 6 and 10).

For what concerns the energy we have seen that the energy of the sole magnetic field (eq. 27' and 28) balances the hidden energy.

But consider now the work made by the external accelerating forces, at least the part that counterpoises the force due to the induced field $E_i$. This field comes only from the magnetic field variation so it is natural to expect that it gives account only of the magnetic energy:

$$L_{E_i} = -\int 2qE_i v dt = \mu_0 sA\sigma^2 \int \dot{v}v dt = \frac{\mu_0}{2} sA\sigma^2 v^2 = E_m = 2T_e \tag{33}$$

This work comes only from the internal reaction forces responsible of the hidden energy.

The rest of the work made by the external accelerating forces goes in the increased energy of the electric field, because the length contraction gives rise to an increased density of the electric charges on the two plates of the capacitor. And this requires an amount $T_e$ of energy, which is in fact the kinetic energy (see eq.s 27 and 27').

### III. CAPACITOR MOVING IN DIRECTION PERPENDICULAR TO THE PLATES

In this case the mechanical supports that contrast the electrostatic attraction of the two plates are under compressing stress in the translation direction. In this case the hidden momentum is in the same direction of the velocity. In fact the energy current $G$ goes from the back side to the front side. Le us suppose that $A$ represents the overall cross section of the supports (in principle the dielectric itself), then we have

$$G = \frac{1}{2\varepsilon_0} D^2 Av' = -\frac{1}{2\varepsilon_0} D^2 Av \tag{34}$$

and the hidden momentum turns out to be

$$p_h = \frac{Gs}{c^2} = \frac{\mu_0}{2}\sigma^2 Asv' = -\frac{\mu_0}{2}\sigma^2 Asv \tag{35}$$

The kinetic linear momentum of the electric field is

$$p_e = m_{0e}v' = \frac{\mu_0}{2}\sigma^2 Asv' = -\frac{\mu_0}{2}\sigma^2 Asv \tag{36}$$

Because of the fact that $B=0$ in these conditions the electromagnetic momentum vanishes. As before we see that the balance equation is verified, i. e.

$$p_e = q_{em} + p_h \tag{37}$$

And what about the energy? The positive work made by the electric forces on the support in the time

$$\Delta t = \frac{vs}{c^2} \tag{38}$$

is given by

$$\Delta L = \frac{\sigma^2}{2\varepsilon_0} A \cdot v \cdot \frac{vs}{c^2} = \frac{\mu_0}{2}\sigma^2 v^2 As = 2T_e \tag{39}$$

so that there is an energy gain $2T_e$ for the hidden energy.

The electromagnetic energy in the $S'$ reference frame is

$$\mathsf{E}'_{em} = \frac{1}{2} s' A' \frac{D'^2}{\varepsilon_0} \tag{40}$$

being now $D' = D, A' = A$ and $s' = s/\gamma$

Hence the change in electromagnetic energy is

$$\Delta \mathsf{E}_{em} = \mathsf{E}'_{em} - \mathsf{E}_{em} = \frac{1}{\gamma} c^2 m_{0e} - c^2 m_{0e} = \left(1 - \frac{1}{2}\beta^2 - 1 + ...\right) m_{0e} c^2 = -\frac{1}{2} m_{0e} v^2 = -T_e \tag{41}$$

As a result we see that the same energy balance equation (29) between the different terms still holds. But in this case it is the hidden energy that acquires an amount $2T_e$ while the electromagnetic field loses an amount $T_e$.

## IV. ACKNOWLEDGMENTS

The author is grateful to Prof. Roberto Coisson for helpful discussions and various suggestions for the text.

\* NOTE 1 – The present analysis is made in the approximation of small velocities keeping only the lowest order term. Thus many of the results here discussed do show linear dependence on the velocity, mostly when the linear momentum is involved. Effects involving energy exchange show instead square dependence On the other hand it is peculiar of electromagnetism to be a clamorous manifestation of relativity at the typical velocities of every day life

\*\* NOTE 2 – Hereafter vector quantities parallel to x-axis are mostly indicated through their x-component, when it is clear from the context.

\*\*\* NOTE 3 - It is worth noting that the hidden momentum can also be connected directly with another important relativistic principle, the relativity of the simultaneity [12]. In fact we can also calculate it by supposing that the tension $\tau$ is activated at a certain instant $t$ in the rest system $S$ [2]. But in the system $S'$ the tension on the back side of the capacitor plates appears before that of the front side by a time interval $\Delta t$ deduced from Lorentz transformation $t' = \gamma(t - vx/c^2)$. Hence the back side receives an uncompensated impulse given by

$$p_h = 2l\tau \cdot \Delta t = 2l \frac{\sigma^2 s}{4\varepsilon_0} \cdot \frac{va}{c^2} = \frac{\mu_0}{2} sA\sigma^2 v \tag{5'}$$